\documentclass[sigconf]{acmart}

\usepackage{booktabs} % For formal tables
\usepackage{multirow}
\usepackage{mathtools}
\usepackage{subcaption}
\usepackage{float}
\usepackage{bm}
\usepackage{enumitem}

\setcopyright{rightsretained}
\usepackage{tabularx}

\copyrightyear{2018} 
\acmYear{2018} 
\setcopyright{acmcopyright}
\acmConference[RecSys '18]{Twelfth ACM Conference on Recommender Systems}{October 2--7, 2018}{Vancouver, BC, Canada}
\acmBooktitle{Twelfth ACM Conference on Recommender Systems (RecSys '18), October 2--7, 2018, Vancouver, BC, Canada}
\acmPrice{15.00}
\acmDOI{10.1145/3240323.3240394}
\acmISBN{978-1-4503-5901-6/18/10}

\acmPrice{15.00}

\begin{document}
%\title{CPRec: Consumer and Producer based Recommendation for Web 2.0 Applications}
% \title[CPRec: Consumer and Producer based Recommendation]{CPRec: Learning Consumer and Producer Embeddings for Recommendation on User-generated Content based Applications}
\title[CPRec: Consumer and Producer based Recommendation]{Learning Consumer and Producer Embeddings for User-Generated Content Recommendation}

\author{Wang-Cheng Kang}
% \authornote{Dr.~Trovato insisted his name be first.}
\affiliation{%
  \institution{UC San Diego}
  %\streetaddress{P.O. Box 1212}
  %state{CA}
  %country{USA}
}
\email{wckang@ucsd.edu}

\author{Julian McAuley}
% \authornote{Dr.~Trovato insisted his name be first.}
\affiliation{%
  \institution{UC San Diego}
  %\streetaddress{P.O. Box 1212}
  %state{CA}
  %country{USA}
}
\email{jmcauley@ucsd.edu}

%\renewcommand{\shortauthors}{W.-C. Kang et al.}

\begin{abstract}
User-Generated Content (UGC) is at the core of 
%Web 2.0
web
applications
%, like ebay, twitter, Flickr, etc.,
where users can both produce and
%browse 
consume
content. This differs from 
traditional
e-Commerce
domains 
%JULIAN: Haven't defined b2c. But I'd avoid defining too much in the abstract as it sounds jargony
%like b2c commerce, Movies, books, etc., 
where 
%the 
%JULIAN: 2nd sentence is fine but a bit redundant as it's already implied by the first
content producers and consumers are usually from two separate groups. In this work, we propose a method \emph{CPRec}
(consumer and producer based recommendation), 
for recommending content on UGC-based 
%applications. 
platforms.
Specifically, we learn a core embedding for each user and two transformation matrices to project the user's core embedding into  two `role' embeddings (i.e., a producer 
%embedding and consumer embedding). 
and consumer role).
We model each interaction by the 
%interactions 
ternary relation
between the consumer, the consumed item, and its producer. Empirical studies on two 
%popular
large-scale
UGC applications show 
that
our method 
%can 
outperforms standard collaborative filtering methods as well as 
recent
methods 
%using the 
that model
producer information via item features.
\end{abstract}




\maketitle

\newcommand{\xhdr}[1]{\vspace{0.8mm}\noindent{{\bf #1}}}

\section{Introduction}

\begin{figure}[t]
\centering
\includegraphics[width=0.8\linewidth]{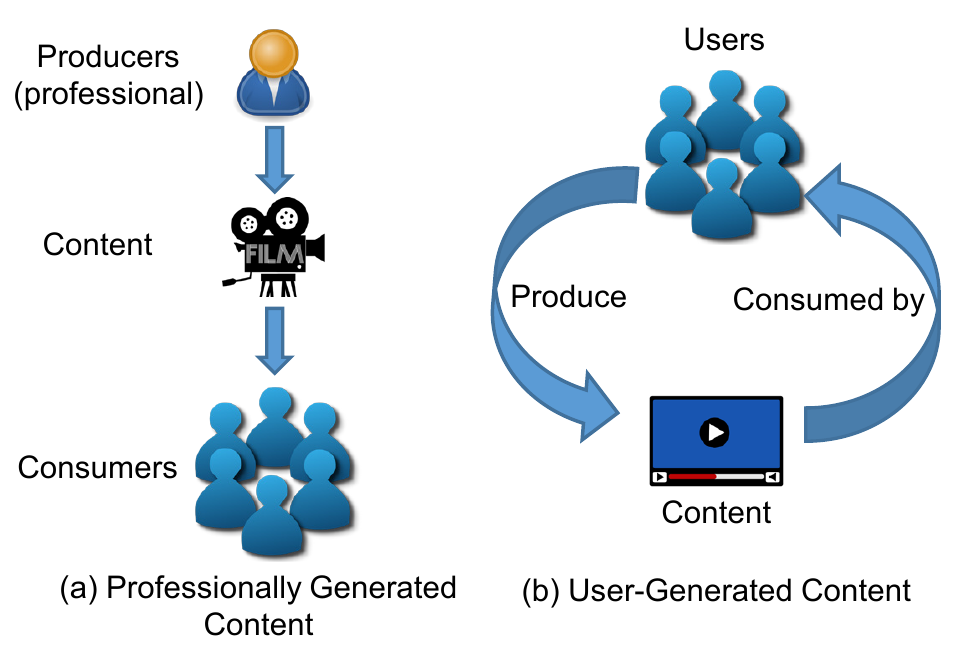}
\caption{An illustration of the production and consumption processes of professionally generated content versus UGC. In centralized domains, producers and consumers form two separate groups. In contrast, users simultaneously perform the two roles (i.e., being a `prosumer') in UGC platforms.}
%JULIAN: Neet to change "contents" to "content" in caption.
%JULIAN: This figure borders on being too obvious. It's not explaining a concept that's really difficult to grasp, but rather creates the impression that your idea is *very* simple. It may turn readers off.
\label{fig:fig1}
\end{figure}

We consider the problem of providing recommendations on user-generated content (UGC) communities.
Unlike traditional domains for recommendation, UGC communities form a `closed loop' between users and items (Figure \ref{fig:fig1}).
%Except the large amount and the lack of side information, the flow of UGC is a close loop betweens users and items (as showed in Figure 1).
That is,
%to say, 
each user can have two roles: a consumer and a producer. However, conventional recommender systems in centralized domains (e.g.~Amazon, Netflix) only focus on
%user's 
users'
consumption behavior (e.g.~clicks, purchases, 
%watches, 
views,
etc.),
i.e., the relationship between the consumer and the item.
%. Furthermore, each consumption 
%behavior of 
%in
However for UGC applications we have additional dynamics to model, namely the ternary relationship between the consumer, the item, and the item's producer (who is in turn also a user).
%UGCs is essentially related to the consumer, the item, and its producer, while exiting approaches usually models a consumer-item interaction. Hence, building a appropriate model for recommending user-generated content while simultaneously consider the consumer and producer roles of users is the main purpose of this paper.

Our main goal in this paper is to design models for recommendation that explicitly consider this ternary dynamic. Specifically, we propose a method \emph{CPRec} (consumer and producer based recommendation), which learns two role embeddings derived from the same core embedding via two projection matrices. We model users' consumption behavior by the summation of
%JULIAN: Avoid
%his
her preference toward the item and her `appreciation' toward the item's producer. We compare our method against various baselines on 
%JULIAN: Can probably be assumed that it's real data
%real 
two
UGC platforms (\emph{Pinterest} and \emph{Reddit}).
%JULIAN: Can safely be assumed
%Our results show that \emph{CPRec} can generate more accurate recommendation in terms of ranking performance.
%JULIAN: Describe your solution...

%prosumer~\cite{giurgiu2008prosumer}

\section{Related Work}
%Recommender systems; ownership-aware recsys; Social Recommendation
\xhdr{Recommender Systems:}
Recommender systems focus on modeling the compatibility between users and items, based on historical feedback (e.g.~clicks, purchases, likes). Matrix Factorization (MF) methods seek to uncover latent dimensions to represent 
%user's
users' preferences and
%items
items'
%property, 
properties,
and estimate 
%the 
user-item interactions through the inner product between the user embedding and item embedding \cite{Handbook,korenSurvey}. 
%The 
User feedback can be \emph{explicit} (e.g.~ratings) or \emph{implicit} (e.g.~clicks, purchases, comments)~\cite{WRMF,rendle2009bpr}. 
%Utilizing 
Modeling
implicit feedback 
%could 
can
%JULIAN: Confusing, this paper isn't really about "interpretability"
be challenging %in how to interpret 
due to the ambiguity of interpreting `non-observed' (e.g.~non-purchased) data.
Recently, \emph{point-wise}~\cite{WRMF} and 
%JULIAN: I don't think the insight is about adapting MF, it's about making use of ranking losses. The use of MF isn't particularly significant here.
\emph{pairwise}~\cite{rendle2009bpr} methods are proposed to solve such challenges.
%have successfully adapted MF to address such challenges.

%JULIAN: First part of related work seems more relevant to a paper about implicit feedback rather than anything specific to your contribution

\emph{Point-wise} methods assume non-observed feedback to be inherently negative, and model the problem via regression,
%which for each user-item pair predicts an affinity score and then ranks items %accordingly.
either by assigning 
`confidence levels' to 
%positive and non-observed 
feedback \cite{WRMF}, or by sampling non-observed feedback as negative instances \cite{OCCF}.

%In contrast to point-wise methods,
\emph{Pairwise} methods are based on a weaker but possibly more realistic assumption that positive feedback must only be `more preferable' than non-observed feedback.
Such methods directly optimize the ranking (in terms of the AUC) of the feedback and are to our knowledge state-of-the-art for implicit feedback datasets. In particular, Bayesian Personalized Ranking (BPR), has
%experimentally show that BPR-MF (i.e., with MF as the underlying predictor)
experimentally been shown to 
outperform a variety of competitive baselines \cite{rendle2009bpr}.

%JULIAN: Sentence is a bit out of place. You might explain the different facets of related work at the start of the section
In this work, we treat users' consuming behaviors as implicit feedback, and seek to optimize their personalized pairwise ranking.

\xhdr{Ownership-Aware Recommendation:}
%Except the 
Other than
standard MF approaches for estimating user-item interactions,
%JULIAN: FMs don't really fit in this section
Factorization Machines (FMs)~\cite{FM} provide a generic factorization approach that can easily incorporate 
%with side 
side
%features 
information
of users and items. 
In our case,
%The
producer information 
%we used 
can be 
%also 
viewed as an item 
%side 
%JULIAN: Weird to put this in related work, it seems more to belong in the model section
feature (via a one-hot representation), that can be used by FMs. Recently, Vista~\cite{Vista} was proposed for artistic recommendation with ownership information. Though the two methods can make use of ownership information, they do not specifically model the two roles (consumer and producer) of each user as we do. We discuss these methods in more detail in the next section, and compare our method against them empirically.

\xhdr{Socially-Aware Recommendation:} 
Leveraging social networks can help us understand user-user relationships and improve the performance of item recommendation~\cite{DBLP:conf/cikm/ZhaoMK14,DBLP:conf/wsdm/MaZLLK11,DBLP:conf/wsdm/Krohn-GrimbergheDFS12,DBLP:conf/recsys/RafailidisC17}. The social network is usually based on friendship or `trust' relationships. A typical assumption in socially-aware methods is that users prefer to follow their friends' behaviors since they may share common interests. However, in our problem setting, there is not such an explicit social network between users, though our model tries to uncover a implicit `follow' relationships from users' consumption and production patterns.

\xhdr{Heterogeneous User Feedback Modeling:} Unlike conventional recommendation methods that only consider a single type of feedback, there is a line of work that seeks to model users' different types of behavior. For example, DualRec~\cite{DBLP:conf/cikm/WangTL15} considers a reviewer role and a rater role (rates the helpfulness of reviews).
%of users.
A recent method SPTF \cite{DBLP:conf/icdm/YinCSWWN17} is a more general approach to jointly model different types of user behavior (e.g.~click, add-to-cart, purchase) via probabilistic tensor factorization. However, these methods typically focus on modeling users' consumption behaviors, whereas we consider both consumer and producer behaviors in UGC platforms.

\section{CPRec: Consumer and producer based recommendation}

\begin{table}[t]
\caption{Notation. \label{tab:notation}}
%JULIAN: Useful package
\begin{tabularx}{\linewidth}{lX}

\toprule
Notation&Explanation\\
\midrule
$\mathcal{U},\mathcal{I}$ & user and item set\\
$\mathcal{I}_u^+$ & positive item set for user $u$\\
$p_i\in \mathcal{U}$ & the producer of item $i$\\
$\widehat{x}_{ui}\in\mathbb{R}$ & predicted score user $u$ gives to item $i$ \\
$K\in \mathbb{N}$ & latent factor dimensionality\\
$\bm{\gamma}_u\in \mathbb{R}^K$	& core embedding for user $u$\\
$\mathbf{W}^c_u,\mathbf{W}^p_u \in \mathbb{R}^{K\times K}$	& role transformation matrices (consumer and producer)\\
$\bm{\gamma}_i\in \mathbb{R}^K$	& item $i$'s embedding\\
$C\subset\mathcal{U}$		&	set of all consumers\\
$P\subset\mathcal{U}$		&	set of all producers\\
$\mathit{PS}=C\cap P$	&	set of all prosumers\\
\bottomrule
\end{tabularx}
\end{table}

\subsection{Problem Description}

We consider a system that
%making 
makes
%JULIAN: Suggests content-aware
%content 
recommendations on
%JULIAN: Not sure if this term is meaningful?
%user-centric
%JULIAN: Jargon, doesn't belong in an academic paper
%Web 2.0
UGC applications with implicit feedback (e.g.~click, comment, retweet, etc.). Since there is no observed negative feedback, the goal is to rank items such that 'observed' items 
should be ranked
higher than non-observed items. A critical property in this domain is that 
%the 
users not only provide 
%consume 
feedback on items, but also that all 
%the
items 
%on the platform 
are created by 
the users
themselves. 
%That is to say each user performs two roles: consumers and producers.
That is, each user assumes both the role of a consumer and producer.
%Specifically, 
We use $\mathcal{U}$ and $\mathcal{I}$ to represent the set of users and items (respectively).
%JULIAN: Try to avoid too much unnecessary language
%in the application. 
For each user $u$, we use $\mathcal{I}^+_u$ to denote all items toward which 
she
%her 
has provided positive feedback.
Finally,
each item $i\in\mathcal{I}$ is produced by the user $p_i\in\mathcal{U}$. 

We define the sets $C\subset\mathcal{U}$ and $P\subset\mathcal{U}$ to represent consumers (who provided any feedback) and producers (who created any item). In addition to $C\cup P=\mathcal{U}$, we have 
%JULIAN: Remember to use mathit for multi-letter variables
$\mathit{PS}=C\cap P$ for representing `prosumers' who both created and consumed items. The ratio $|\mathit{PS}|/|\mathcal{U}|$ is critical to identify how the groups of consumers and producers 
% are overlapped.
overlap. Table \ref{tab:notation} summarizes our notation.
%In 
%Web 2.0 applications based on user-generated content,
%JULIAN: Seems not interesting, as having this ratio be non-zero is precisely the *definition* of UGCs
% UGC applications
% this ratio is often relatively high, for example, many people both sell and buy items on \emph{ebay}, or both upload and save images on \emph{Pinterest}. However, this is not the case in domains like b2c e-commerce (\emph{Amazon}), movies (\emph{Netflix}), books (\emph{Goodreads}). The producers (i.e., manufacturers, directors, writers) and the consumers in these domains are totally separated. (i.e., $|PS|/|\mathcal{U}|=0$).

\subsection{The CPRec Model}

Biased matrix factorization is widely used as
an
underlying preference predictor in recommendation problems \cite{rendle2009bpr,koren2008factorization}. Specifically, it models user-item interactions via bias terms and 
%a
an
inner product between the latent vectors of the user and item:
\[\widehat{x}_{ui}=\alpha+\beta_{u}+\beta_{i}+\langle\bm{\gamma}_u,\bm{\gamma}_i\rangle\]
Though this has shown strong performance in modeling user-item interactions, 
it does not fully model ternary interactions between
%in domains we consider, each interaction is essentially related to 
the consumer $u$, the item $i$, and its producer $p_i$. Hence we propose a model to capture this interaction by
%factorizing 
factorization
into two parts:
\[\widehat{x}_{ui}=\alpha+\beta_{u}+\beta_{i}+\underbrace{\langle\bm{\gamma}_u^c,\bm{\gamma}_i\rangle}_{\mathclap{\text{consumer-item preference}}}+\overbrace{\langle\bm{\gamma}_u^c,\bm{\gamma}_{p_i}^p\rangle}^{\mathclap{\text{consumer-producer appreciation}}}.\]
We introduce two embeddings ($\bm{\gamma}_u^c$ and $\bm{\gamma}_u^p$) to represent the user $u$'s two roles (consumer and producer). This is mainly because: (1) The 
%JULIAN: Wrong quote symbols
%'follow' 
`follow'
relationship between users is asymmetric, which cannot be modeled by the inner product of homogeneous embeddings (i.e., using the same user embedding for her two roles); 2) Users may
%show
exhibit
different behavior when they play different roles.
%JULIAN: Fairly obvious, I don't think you need to explain this fairly simple idea in so much detail
%and heterogeneous embeddings are capable of modeling this due to its higher flexibility.
Ultimately, we model user $u$'s consumer embedding and producer embedding as being derived from a single core embedding $\bm{\gamma}_u$ and two transformation matrices:
\[\bm{\gamma}_u^c=\mathbf{W}^c\bm{\gamma}_u,\qquad\bm{\gamma}_u^p=\mathbf{W}^p\bm{\gamma}_u\]
where $\mathbf{W}^c,\mathbf{W}^p\in\mathbb{R}^{K\times K}$. That is to say, we use two projection matrices to project a user's core embedding into her two role embeddings. The advantage here is we only introduce $2K^2$ new parameters (compared to standard MF) to achieve asymmetric embeddings, which is helpful to avoid overfitting.

\subsection{Learning}

Based on the proposed preference predictor, we adopt a Bayesian Personalized Ranking (BPR) framework \cite{rendle2009bpr} to learn all parameters. The goal of BPR is to approximately optimize the AUC of ranking observed feedback for each user. Specifically, we consider the triplets $(u,i,j)\in\mathcal{D}$, where:
\[\mathcal{D}=\{(u,i,j)|u\in\mathcal{U}\wedge i\in\mathcal{I}_u^+\wedge j\in\mathcal{I}\setminus \mathcal{I}_u^+\}.\]

Here $i\in\mathcal{I}_u^+$ is an item about which the user $u$ has provided feedback, whereas $j\in\mathcal{I}\setminus \mathcal{I}_u^+$ is one about which they have not. Thus intuitively, for a user $u$, the predictor should assign a larger preference score
to item $i$ than item $j$. Hence BPR defines the difference between preference scores by
\[\widehat{x}_{uij}=\widehat{x}_{ui}-\widehat{x}_{uj}.\]
Note that the global bias term $\alpha$ and user bias term $\beta_u$ are naturally canceled in $\widehat{x}_{uij}$. We seek to optimize the ranking by maximizing the posterior
\begin{equation*}
\begin{split}
\ln p(\Theta|\mathcal{D})& \propto \ln\prod_{(u,i,j)\in \mathcal{D}} \sigma (\widehat{x}_{uij})p(\Theta)\\
&=\sum_{(u,i,j)\in \mathcal{D}} \ln\sigma (\widehat{x}_{uij})-\lambda_\Theta\|\Theta\|^2,
\end{split}
\end{equation*}
where $\sigma(\cdot)$ is the sigmoid function, $\Theta=\{\bm{\gamma}_u,\bm{\gamma}_i,\beta_i,\mathbf{W}^c,\mathbf{W}^p\}$ includes all model parameters, and $\lambda_\Theta$ is a regularization hyperparameter. We adopt 
the
Adam optimizer~\cite{DBLP:journals/corr/KingmaB14}, a variant of stochastic gradient descent with adaptive estimation of moments, to learn all variables. 

\subsection{Discussion of Related Methods}

Vista~\cite{Vista} is a recent method for artistic recommendation (applied to data from \emph{behance.net}), which leverages
%the 
ownership information.\footnote{Vista also considers visual and temporal information, however they are not the focus of this paper.
%JULIAN: Not a good place to discuss future work, before you've even shown results. It will deflate readers' excitement about what you've done
%In the future work, we consider extending our model to incorporate more features
}
Vista models user-item interactions via:
\[\widehat{x}_{ui}=\langle\bm{\gamma}^{(1)}_u,\bm{\gamma}_i\rangle+\langle\bm{\gamma}^{(2)}_u,\bm{\gamma}^{(2)}_{p_i}\rangle.\]
One major difference is that Vista uses the inner product of symmetric embeddings to model relationships between users. However, as we stated before, the ``follow'' relationship between users may be asymmetric in the domains we consider. Hence our model 
%JULIAN: Overreaching a bit rather than just letting your results do the talking
%is
may be
more suitable for capturing the relationship between users.

Another way to use 
%the information of producer is 
producer information is to
view it as a categorical item feature. Factorization Machines (FMs) provide a generic factorization method by modeling 
%the 
interactions between users, items, and their features. Here, we assign each item $i$ a one-hot feature $\mathbf{f}_i\in\{0,1\}^{|\mathcal{U}|}$, and $\mathbf{f}_{p_i}=1$. The (second-order) estimator of FMs is given by:
\[\widehat{x}_{ui}=\beta_u+\beta_i+\beta_{p_i}+\langle\bm{\gamma}^{(1)}_u,\bm{\gamma}_i\rangle+\langle\bm{\gamma}^{(1)}_u,\bm{\gamma}^{(2)}_{p_i}\rangle+\langle\bm{\gamma}_i,\bm{\gamma}^{(2)}_{p_i}\rangle.\]
FMs can capture asymmetric user relationships since they make use of two 
%JULIAN: Informal
%totally 
different embeddings (i.e., $\bm{\gamma}^{(1)}_u,\bm{\gamma}^{(2)}_{p_i}$) to model their interaction. However, 
%JULIAN: Avoid too many contractions (informal)
%it's
it is not aware
%of that 
that
$\bm{\gamma}^{(1)}_u$ and $\bm{\gamma}^{(2)}_u$ are 
%essentially 
the same 
%person. 
individual.
We argue 
that
the two embeddings should be connected since
%user's 
users'
behaviors under the two roles should be related (yet different).

%There are other social recommendation methods, 
Other than the above, there are a few methods that focus on `socially-aware' recommendation, such as
Social-BPR~\cite{DBLP:conf/cikm/ZhaoMK14}. However, our problem setting differs from that 
%in 
of
social recommendation as we don't have
%a 
an
explicit social network of users, 
%and
but rather
we focus on modeling interactions which 
%is 
are
related to $(u,i,p_i)$ where $p_i$ is the
%unique
creator of item $i$.

\section{Empirical Studies}
\subsection{Dataset}

We consider two public datasets from UGC-based
%JULIAN: I really suggest you avoid saying this. These types of terms are badly dated and will prevent people from taking your work seriously.
%WC: Got it.
%Web 2.0 
applications:
\begin{itemize}[leftmargin=*]
\item \textbf{Pinterest} is a content discovery application based mainly around images. People can browse, upload (pin), like, and save (repin) images. We use the dataset crawled by \cite{icwsm2013zhong},\footnote{\url{https://nms.kcl.ac.uk/nishanth.sastry/projects/cd-gain/dataset.html}} which includes 0.89M users, 2.4M images and 56M actions (`save' and `like') in January 2013 . We 
%take 
treat
both ``like'' and ``repin'' actions as implicit feedback. Each item 
%associates 
is associated
with
%a 
an
uploader.
\item \textbf{Reddit} is a discussion website which covers a variety of topics including news, science, movies, etc.
%People 
Users
can 
%share 
submit
content
%(called a submission) 
and comment on submissions. We use a dataset
%JULIAN: Too informal
%from here, 
which includes all 
%the 
submissions and comments on Reddit in March 2017.\footnote{\url{https://redd.it/6607j2}} Specifically, the dataset includes 1.3M users, 9.6M submissions 48M comments. We view each submission as an item, and 
%take the action of comment 
commenting actions
as implicit feedback. Each submission 
%associates with an 
is associated with a single
author.
\end{itemize}

\begin{figure}[t]
\centering
\begin{subfigure}[b]{0.49\linewidth}
\includegraphics[width=\linewidth]{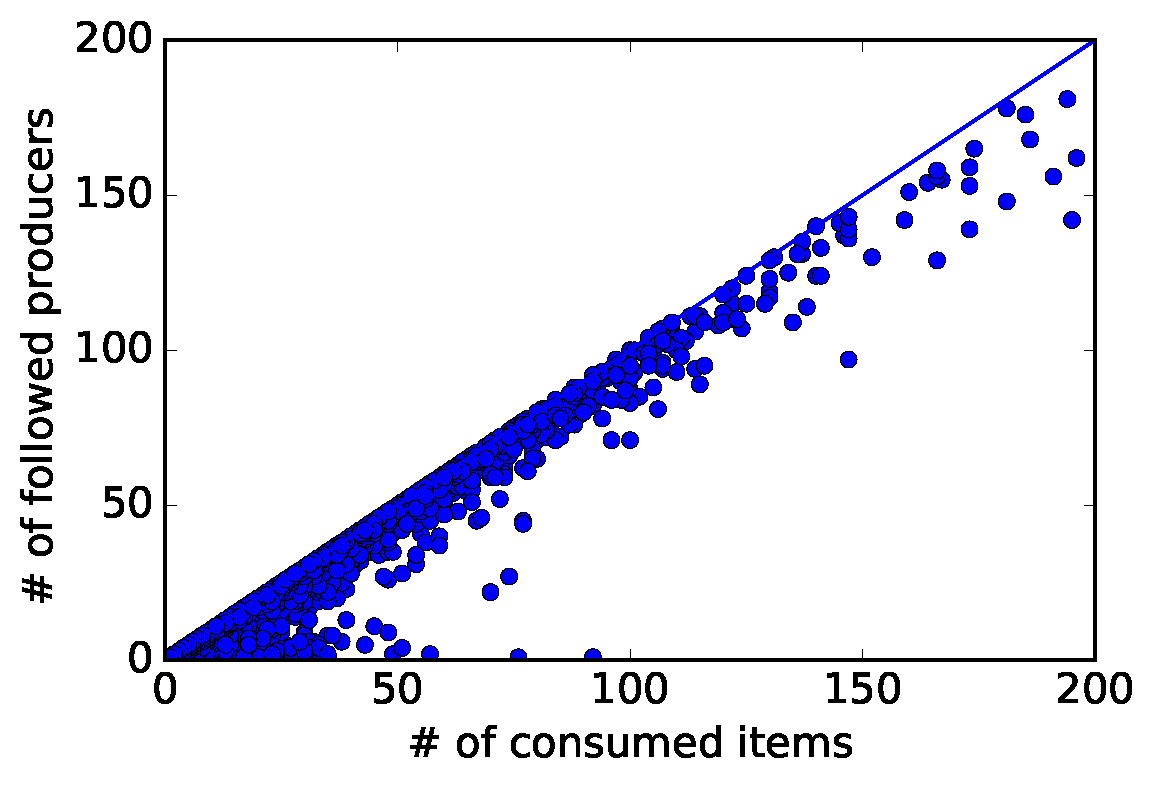}
\subcaption{Pinterest}
\end{subfigure}
\begin{subfigure}[b]{0.49\linewidth}
\includegraphics[width=\linewidth]{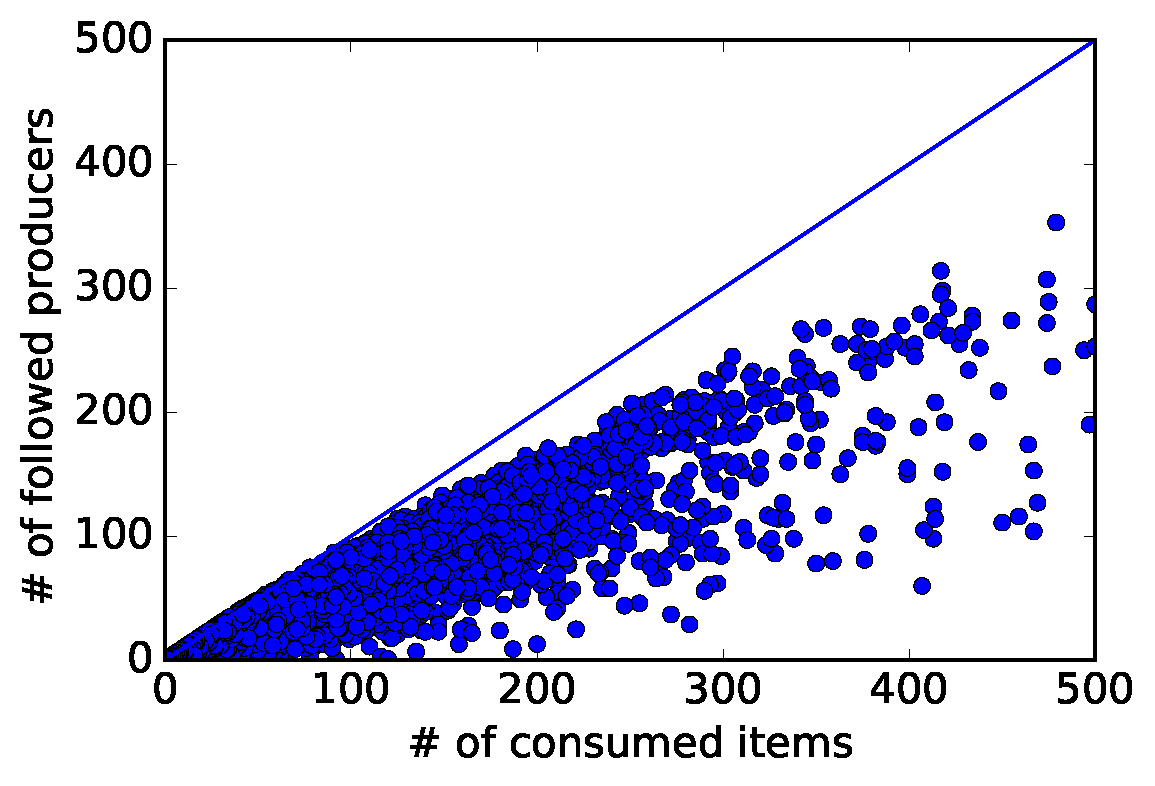}
\subcaption{Reddit}
\end{subfigure}
\caption{Number of followed producers vs.~number of consumed items. Each point represents a user. If user $u$ consumed $N$ items that are created from $N$ different producers, her point in the figure will lie on the line. This shows
%, 
that
Reddit users are more inclined to
%JULIAN: Wrong word
%duplicated 
% highly active
%JULIAN: Has a different meaning
%repetitive
repeatedly follow the same
producer.}
\label{fig:fig2}
\end{figure}

For data preprocessing, we discard inactive users and items which have
%less
fewer
than 10 associated actions. For each user, we randomly withhold one action for validation $\mathcal{V}_u$, and another for testing $\mathcal{T}_u$. All remaining items are used for training $\mathcal{P}_u$, and we always tune models via our validation set and report the performance on 
the
test set. Table \ref{exp:dataset} lists statistics of our datasets. Figure~\ref{fig:fig2} shows 
users
%repetitively 
repeatedly
following appreciated producers. 
%the
%JULIAN: Doesn't make sense. I guess duplicated is the wrong word.
%following duplicated producers. 
%following behavior as a function of producer activity.
We also considered the \emph{behance} dataset used in \cite{Vista}, however, 
%it's not suitable 
it proved unsuitable
for our problem since its prosumer ratio is almost zero, which indicates
%its producers and consumers are basically from two separated groups.
that most of its users only have a single role (consumer or producer).

\begin{table}[t]
\centering
\caption{Dataset statistics (after preprocessing) \label{exp:dataset}}
%\setlength{\tabcolsep}{3pt}
%JULIAN: Right align numerical columns so that readers can compare in terms of significant figures
\begin{tabular}{lrr}
\toprule
 & Pinterest & Reddit\\ \midrule
\#users ($|\mathcal{U}|$) & 134,747   & 52,654   \\
\#items ($|\mathcal{I}|$)  & 201,792   & 336,743   \\
\#actions ($\sum_{u\in\mathcal{U}}|\mathcal{I}^+_u|$)    & 690,506   & 1,786,032  \\
%\#consuming actions/user	&	5.12	&	33.92\\
%\#producing actions/user	&	1.50	&	6.16\\
consumer ratio ($|C|/|\mathcal{U}|$)        & 93.65\%   & 99.60\%  \\
producer ratio ($|P|/|\mathcal{U}|$)       & 80.76\%   & 87.24\%  \\ 
prosumer ratio ($|\mathit{PS}|/|\mathcal{U}|$)       &74.42\%    & 86.85\% \\
\bottomrule
\end{tabular}
\end{table}

\subsection{Baselines}
We compare our method with standard recommendation methods as well as methods 
%utilizing the information about producers.
that make use of producer information.
\begin{itemize}[leftmargin=*]
\item \textbf{PopRec} is a straightforward baseline which ranks items according to their popularity.
\item \textbf{Bayesian Personalized Ranking (BPR)}~\cite{rendle2009bpr} is a state-of-the-art recommendation method for implicit feedback. We use biased matrix factorization as the underlying predictor ($\widehat{x}_{ui}=\beta_i+\langle\bm{\gamma}_u,\bm{\gamma}_i\rangle$).
\item \textbf{Factorization Machines (FMs)}~\cite{FM} provide a
generic factorization approach that can be used to model interactions between users, items, and their features. We use a one-hot encoding to represent the producer of each item.
\item \textbf{Visually, Socially, and Temporally-Aware Recommendation (Vista)}~\cite{Vista} is a recent method for artistic recommendation. We use a reduced model which only considers 
%the
ownership information.
\end{itemize}

%JULIAN: I don't think this paragraph adds anything
% And our method \textbf{CPRec}, which considers user-item interaction as well as consumer-producer relationship, and learns two role embeddings derived from the same user embedding.

These baselines are intended to show (a) the importance
of learning personalized notions of compatibility (MF methods vs.~PopRec); (b) the effect of modeling consumer-producer interactions for 
%user-generated content based 
UGC
%Web 2.0 
applications (FM/Vista/CPRec vs.~BPR); and (c) the improvement gained by our consumer/producer embedding approach (CPRec vs.~FM/Vista).

We implemented all MF methods using \emph{Tensorflow}~\cite{tensorflow}. For fair comparison, we train all 
%the 
methods using the BPR loss and Adam optimizer~\cite{DBLP:journals/corr/KingmaB14}. 
%we 
We
tune hyperparameters via grid search on a validation set, with a regularizer selected from \{0.001, 0.01, 0.1, 1\}. The learning rate is set to 0.01, and the batch size is 10000. The code is available at \url{https://github.com/kang205/CPRec}.

%We will release code at publication time. %We set the dimensionality of latent vector $K=10$ in all the cases.
%JULIAN: Why? Your hyperparameter study seems to show that this is notwhere near the best value of K, and in fact seems to suggest you'd get much better results (both with your own method and compared to others) with a different value.
%WC: Removed. 

\subsection{Recommendation Performance}

We measure the recommendation performance via the 
%metric AUC 
AUC
(Area Under the ROC Curve)~\cite{rendle2009bpr}, which considers the overall ranking:
%Specifically we have:
\[AUC=\frac{1}{|\mathcal{U}|}\sum_{u\in\mathcal{U}} \frac{1}{|\mathcal{D}_u|}\sum_{(i,j)\in\mathcal{D}_u}\xi(\widehat{x}_{ui}>\widehat{x}_{uj}),\]
where $\mathcal{D}_u=\{(i,j)|(u,i)\in\mathcal{T}_u\wedge(u,j)\notin(\mathcal{P}_u\cup\mathcal{V}_u\cup\mathcal{T}_u)\}$ and $\xi(\cdot)$ is an indicator function. Intuitively, AUC is the fraction of times that the `observed' items $i$ are ranked higher than `non-observed' items $j$.

We consider two recommendation target groups: all consumers and cold consumers (who have fewer than five consumed items). We do not consider content recommendation for users who only produced items since we lack ground truth items for evaluation. Table~\ref{tab:recommendation} shows the ranking performance of all methods on the two datasets with latent dimensionality $K=20$ . The overall performance on Reddit is better than that of Pinterest, a possible reason being that Reddit users are more inclined to 
%JULIAN: Wrong word, I tried to rewrite
%follow repetitive 
repeatedly consume content by the same producers (as shown in Figure \ref{fig:fig2}) which makes their behaviors easier to predict.

We can see that FM/Vista/CPRec are 
%JULIAN: Avoid in general -- try to be more precise
%better
more accurate than 
%the standard 
%MF method
BPR (especially for cold users), which shows the importance of considering
%the
producer information in UGC applications. Moreover, \emph{CPRec} outperforms all other baselines in all the settings. Compared to BPR, \emph{CPRec} gains 13.9\% AUC improvement for all users and 18.6\% AUC improvement for cold users on average. Our method also achieves 
a
2.8\% improvement against the strongest baseline on average.

%Another observation is that \emph{CPRec}'s performance for prosumers is slightly better than that for consumers, but this is not the case for FM and Vista. Prosumers are a subset of consumers, and we know more information about prosumers since they both produce and consume items. This shows \emph{CPRec} can handle users with two roles better than FM and Vista, which is a benefit from \emph{CPRec}'s asymmetric consumer/producer embedding approach.

\begin{figure}[htb]
\centering
\begin{subfigure}[b]{\linewidth}
\centering
\includegraphics[width=.99\linewidth]{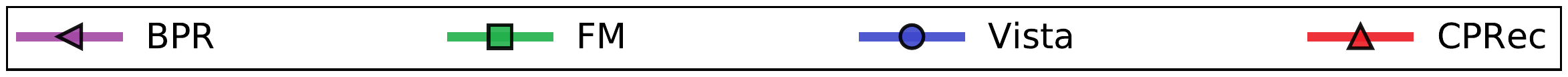}
\end{subfigure}
\begin{subfigure}[b]{0.49\linewidth}
\includegraphics[width=\linewidth]{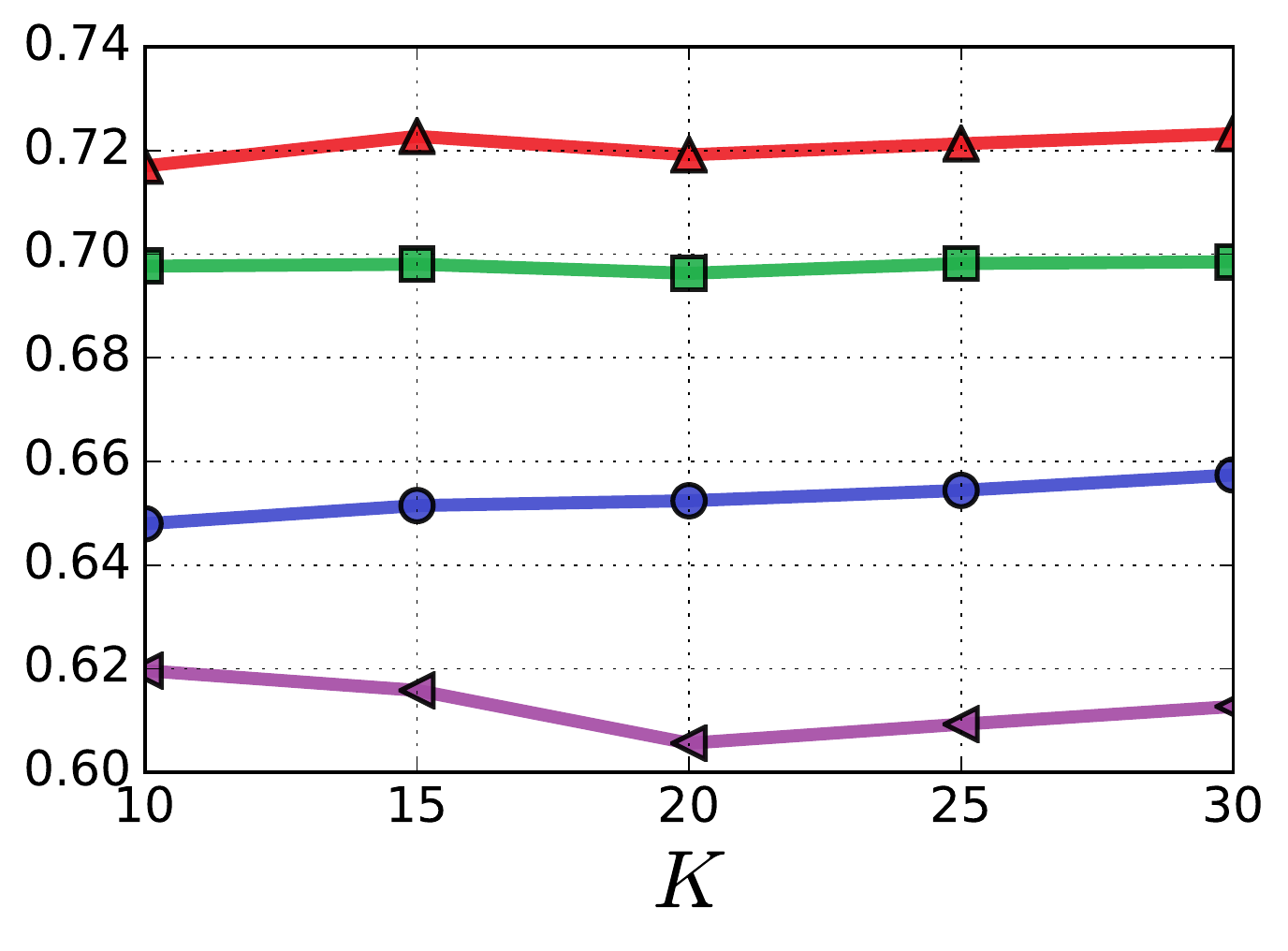}
\subcaption{Pinterest}
\end{subfigure}
\begin{subfigure}[b]{0.49\linewidth}
\includegraphics[width=\linewidth]{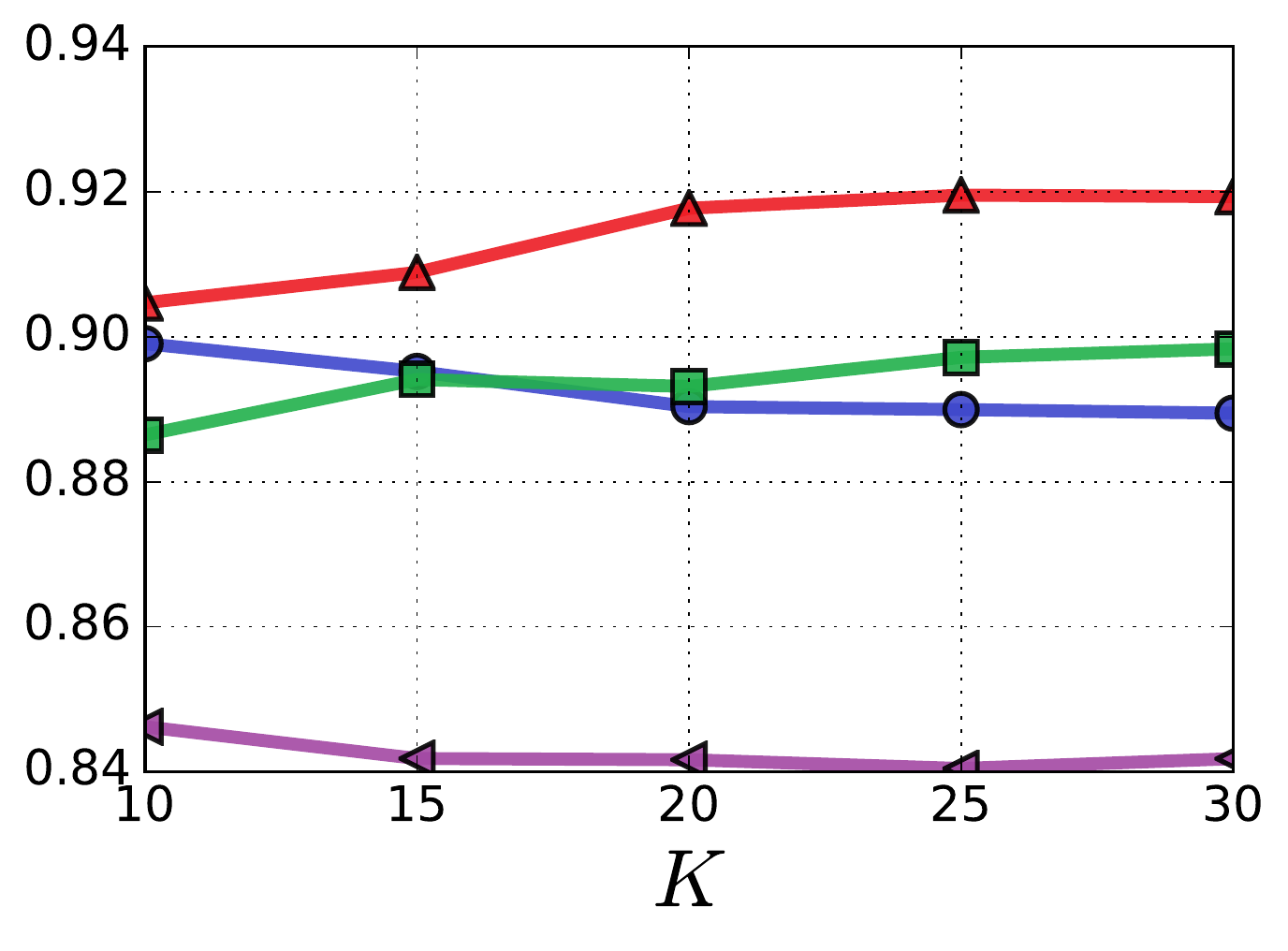}
\subcaption{Reddit}
\end{subfigure}
\caption{Effect of the latent dimensionality $K$. Ranking performance (AUC) of all consumers is shown.}
\label{fig:K}
\end{figure}

To examine the effect of the latent dimensionality $K$ which directly relates to 
%the
model complexity, we plot the ranking performance 
%through
for increasing
%different
$K$ 
%of
for
all MF methods in Figure \ref{fig:K}. We can see that our method \emph{CPRec} is consistently better than the baselines with different $K$. Especially on Reddit, 
%with
as
$K$ increases, the performance gap between \emph{CPRec} and the strongest baseline becomes wider. However, we find that \emph{CPRec} can achieve satisfactory performance with $K=20$ on both datasets.

\begin{table}[htb]
\centering
\caption{Recommendation performance in terms of the AUC with latent dimensionality $K=20$.
\label{tab:recommendation}}
\setlength{\tabcolsep}{3pt}
\begin{tabular}{llccccc}
\toprule
\multirow{2}{*}{Dataset} & \multirow{2}{*}{Target}     & \multirow{2}{*}{\begin{tabular}[c]{@{}c@{}}(a)\\ PopRec\end{tabular}} & \multirow{2}{*}{\begin{tabular}[c]{@{}c@{}}(b)\\ BPR\end{tabular}} & \multirow{2}{*}{\begin{tabular}[c]{@{}c@{}}(c)\\  FM\end{tabular}} & \multirow{2}{*}{\begin{tabular}[c]{@{}c@{}}(d)\\ Vista\end{tabular}} & \multirow{2}{*}{\begin{tabular}[c]{@{}c@{}}(e)\\ CPRec\end{tabular}} \\ \\ \midrule
\multirow{2}{*}{\emph{Pinterest}} & All users  & 0.6125	&	0.6056	&	0.6963	&	0.6524	&	0.7191 \\
%JULIAN: Not sure what was meant by an AUC of zero. Certainly no method should have an AUC of zero, I think it could be misleading to report such a number, rather than leaving the area blank (but maybe this is still something you're planning to fill in?)
%WC: Will fill in.
                                       & Cold users & 0.5993 &	0.5704	&	0.6855	&	0.6316	&	0.6902 \\[1.5mm]                                
\multirow{2}{*}{\emph{Reddit}} & All users  & 0.6397	&	0.8416
&	0.8931	&	0.8903	&	0.9177 \\
                                       & Cold users &	0.5564 & 0.7727	&	0.8668	&	0.8535	&	0.8980 \\                                                        
\bottomrule
\end{tabular}
\vspace{-0.6cm}
\end{table}

\section{Conclusions and Future Work}

In this work, we consider recommendation on user-generated content (UGC) communities, and design a recommendation method \emph{CPRec}, which learns two role embeddings (for consumer and producer roles) derived from the same core user embedding. We model each interaction via a ternary relation between the consumer, the consumed item, and its producer. We analyze the difference between \emph{CPRec} and related methods. Empirically, extensive results on two UGC platforms demonstrate the effectiveness of our method.
In the future, we plan to further investigate the problem of incorporating more context information and side features, including explicit social networks and temporal dynamics.
%For example, take explicit social networks into account (in \emph{CPRec}, we only uncover a latent `follow' relationship between users), model users' temporal dynamics of the two roles, or make the method become content-aware. 
% We also plan to consider other related problems in UGC communities, like producer recommendation and content-producer matching (i.e., anonymous author prediction).

%JULIAN: Note that reference page is free
\bibliographystyle{ACM-Reference-Format}
\bibliography{acmart}


\begin{thebibliography}{17}

%%% the Web address.

\ifx \showCODEN    \undefined \def \showCODEN     #1{\unskip}     \fi
\ifx \showDOI      \undefined \def \showDOI       #1{#1}\fi
\ifx \showISBNx    \undefined \def \showISBNx     #1{\unskip}     \fi
\ifx \showISBNxiii \undefined \def \showISBNxiii  #1{\unskip}     \fi
\ifx \showISSN     \undefined \def \showISSN      #1{\unskip}     \fi
\ifx \showLCCN     \undefined \def \showLCCN      #1{\unskip}     \fi
\ifx \shownote     \undefined \def \shownote      #1{#1}          \fi
\ifx \showarticletitle \undefined \def \showarticletitle #1{#1}   \fi
\ifx \showURL      \undefined \def \showURL       {\relax}        \fi
% invisible to TeX
\providecommand\bibfield[2]{#2}
\providecommand\bibinfo[2]{#2}
\providecommand\natexlab[1]{#1}
\providecommand\showeprint[2][]{arXiv:#2}

\bibitem[\protect\citeauthoryear{Abadi, Barham, Chen, Chen, Davis, Dean, Devin,
  Ghemawat, Irving, Isard, et~al\mbox{.}}{Abadi et~al\mbox{.}}{2016}]%
        {tensorflow}
\bibfield{author}{\bibinfo{person}{Mart{\'\i}n Abadi}, \bibinfo{person}{Paul
  Barham}, \bibinfo{person}{Jianmin Chen}, \bibinfo{person}{Zhifeng Chen},
  \bibinfo{person}{Andy Davis}, \bibinfo{person}{Jeffrey Dean},
  \bibinfo{person}{Matthieu Devin}, \bibinfo{person}{Sanjay Ghemawat},
  \bibinfo{person}{Geoffrey Irving}, \bibinfo{person}{Michael Isard},
  {et~al\mbox{.}}} \bibinfo{year}{2016}\natexlab{}.
\newblock \showarticletitle{TensorFlow: A System for Large-Scale Machine
  Learning.}. In \bibinfo{booktitle}{\emph{OSDI'16}}.
\newblock


\bibitem[\protect\citeauthoryear{He, Fang, Wang, and McAuley}{He
  et~al\mbox{.}}{2016}]%
        {Vista}
\bibfield{author}{\bibinfo{person}{Ruining He}, \bibinfo{person}{Chen Fang},
  \bibinfo{person}{Zhaowen Wang}, {and} \bibinfo{person}{Julian McAuley}.}
  \bibinfo{year}{2016}\natexlab{}.
\newblock \showarticletitle{Vista: A visually, socially, and temporally-aware
  model for artistic recommendation}. In \bibinfo{booktitle}{\emph{RecSys'16}}.
\newblock


\bibitem[\protect\citeauthoryear{Hu, Koren, and Volinsky}{Hu
  et~al\mbox{.}}{2008}]%
        {WRMF}
\bibfield{author}{\bibinfo{person}{Yifan Hu}, \bibinfo{person}{Yehuda Koren},
  {and} \bibinfo{person}{Chris Volinsky}.} \bibinfo{year}{2008}\natexlab{}.
\newblock \showarticletitle{Collaborative filtering for implicit feedback
  datasets}. In \bibinfo{booktitle}{\emph{ICDM'08}}.
\newblock


\bibitem[\protect\citeauthoryear{Kingma and Ba}{Kingma and Ba}{2015}]%
        {DBLP:journals/corr/KingmaB14}
\bibfield{author}{\bibinfo{person}{Diederik~P. Kingma} {and}
  \bibinfo{person}{Jimmy Ba}.} \bibinfo{year}{2015}\natexlab{}.
\newblock \showarticletitle{Adam: {A} Method for Stochastic Optimization}. In
  \bibinfo{booktitle}{\emph{ICLR'15}}.
\newblock


\bibitem[\protect\citeauthoryear{Koren}{Koren}{2008}]%
        {koren2008factorization}
\bibfield{author}{\bibinfo{person}{Yehuda Koren}.}
  \bibinfo{year}{2008}\natexlab{}.
\newblock \showarticletitle{Factorization meets the neighborhood: a
  multifaceted collaborative filtering model}. In
  \bibinfo{booktitle}{\emph{SIGKDD'08}}.
\newblock


\bibitem[\protect\citeauthoryear{Koren and Bell}{Koren and Bell}{2011}]%
        {korenSurvey}
\bibfield{author}{\bibinfo{person}{Yehuda Koren} {and} \bibinfo{person}{Robert
  Bell}.} \bibinfo{year}{2011}\natexlab{}.
\newblock \showarticletitle{Advances in Collaborative Filtering}.
\newblock In \bibinfo{booktitle}{\emph{Recommender Systems Handbook}}.
  \bibinfo{publisher}{Springer}.
\newblock


\bibitem[\protect\citeauthoryear{Krohn{-}Grimberghe, Drumond, Freudenthaler,
  and Schmidt{-}Thieme}{Krohn{-}Grimberghe et~al\mbox{.}}{2012}]%
        {DBLP:conf/wsdm/Krohn-GrimbergheDFS12}
\bibfield{author}{\bibinfo{person}{Artus Krohn{-}Grimberghe},
  \bibinfo{person}{Lucas Drumond}, \bibinfo{person}{Christoph Freudenthaler},
  {and} \bibinfo{person}{Lars Schmidt{-}Thieme}.}
  \bibinfo{year}{2012}\natexlab{}.
\newblock \showarticletitle{Multi-relational matrix factorization using
  bayesian personalized ranking for social network data}. In
  \bibinfo{booktitle}{\emph{WSDM'12}}.
\newblock


\bibitem[\protect\citeauthoryear{Ma, Zhou, Liu, Lyu, and King}{Ma
  et~al\mbox{.}}{2011}]%
        {DBLP:conf/wsdm/MaZLLK11}
\bibfield{author}{\bibinfo{person}{Hao Ma}, \bibinfo{person}{Dengyong Zhou},
  \bibinfo{person}{Chao Liu}, \bibinfo{person}{Michael~R. Lyu}, {and}
  \bibinfo{person}{Irwin King}.} \bibinfo{year}{2011}\natexlab{}.
\newblock \showarticletitle{Recommender systems with social regularization}. In
  \bibinfo{booktitle}{\emph{WSDM'11}}.
\newblock


\bibitem[\protect\citeauthoryear{Pan, Zhou, Cao, Liu, Lukose, Scholz, and
  Yang}{Pan et~al\mbox{.}}{2008}]%
        {OCCF}
\bibfield{author}{\bibinfo{person}{Rong Pan}, \bibinfo{person}{Yunhong Zhou},
  \bibinfo{person}{Bin Cao}, \bibinfo{person}{Nathan~N Liu},
  \bibinfo{person}{Rajan Lukose}, \bibinfo{person}{Martin Scholz}, {and}
  \bibinfo{person}{Qiang Yang}.} \bibinfo{year}{2008}\natexlab{}.
\newblock \showarticletitle{One-class collaborative filtering}. In
  \bibinfo{booktitle}{\emph{ICDM'08}}.
\newblock


\bibitem[\protect\citeauthoryear{Rafailidis and Crestani}{Rafailidis and
  Crestani}{2017}]%
        {DBLP:conf/recsys/RafailidisC17}
\bibfield{author}{\bibinfo{person}{Dimitrios Rafailidis} {and}
  \bibinfo{person}{Fabio Crestani}.} \bibinfo{year}{2017}\natexlab{}.
\newblock \showarticletitle{Learning to Rank with Trust and Distrust in
  Recommender Systems}. In \bibinfo{booktitle}{\emph{RecSys'17}}.
\newblock


\bibitem[\protect\citeauthoryear{Rendle}{Rendle}{2010}]%
        {FM}
\bibfield{author}{\bibinfo{person}{Steffen Rendle}.}
  \bibinfo{year}{2010}\natexlab{}.
\newblock \showarticletitle{Factorization Machines}. In
  \bibinfo{booktitle}{\emph{ICDM'10}}.
\newblock


\bibitem[\protect\citeauthoryear{Rendle, Freudenthaler, Gantner, and
  Schmidt-Thieme}{Rendle et~al\mbox{.}}{2009}]%
        {rendle2009bpr}
\bibfield{author}{\bibinfo{person}{Steffen Rendle}, \bibinfo{person}{Christoph
  Freudenthaler}, \bibinfo{person}{Zeno Gantner}, {and} \bibinfo{person}{Lars
  Schmidt-Thieme}.} \bibinfo{year}{2009}\natexlab{}.
\newblock \showarticletitle{{BPR:} Bayesian personalized ranking from implicit
  feedback}. In \bibinfo{booktitle}{\emph{UAI'09}}.
\newblock


\bibitem[\protect\citeauthoryear{Ricci, Rokach, Shapira, and Kantor}{Ricci
  et~al\mbox{.}}{2011}]%
        {Handbook}
\bibfield{author}{\bibinfo{person}{Francesco Ricci}, \bibinfo{person}{Lior
  Rokach}, \bibinfo{person}{Bracha Shapira}, {and} \bibinfo{person}{Paul
  Kantor}.} \bibinfo{year}{2011}\natexlab{}.
\newblock \bibinfo{booktitle}{\emph{Recommender systems handbook}}.
\newblock \bibinfo{publisher}{Springer US}.
\newblock


\bibitem[\protect\citeauthoryear{Wang, Tang, and Liu}{Wang
  et~al\mbox{.}}{2015}]%
        {DBLP:conf/cikm/WangTL15}
\bibfield{author}{\bibinfo{person}{Suhang Wang}, \bibinfo{person}{Jiliang
  Tang}, {and} \bibinfo{person}{Huan Liu}.} \bibinfo{year}{2015}\natexlab{}.
\newblock \showarticletitle{Toward Dual Roles of Users in Recommender Systems}.
  In \bibinfo{booktitle}{\emph{CIKM'15}}.
\newblock


\bibitem[\protect\citeauthoryear{Yin, Chen, Sun, Wang, Wang, and Nguyen}{Yin
  et~al\mbox{.}}{2017}]%
        {DBLP:conf/icdm/YinCSWWN17}
\bibfield{author}{\bibinfo{person}{Hongzhi Yin}, \bibinfo{person}{Hongxu Chen},
  \bibinfo{person}{Xiaoshuai Sun}, \bibinfo{person}{Hao Wang},
  \bibinfo{person}{Yang Wang}, {and} \bibinfo{person}{Quoc Viet~Hung Nguyen}.}
  \bibinfo{year}{2017}\natexlab{}.
\newblock \showarticletitle{{SPTF:} {A} Scalable Probabilistic Tensor
  Factorization Model for Semantic-Aware Behavior Prediction}. In
  \bibinfo{booktitle}{\emph{ICDM'17}}.
\newblock


\bibitem[\protect\citeauthoryear{Zhao, McAuley, and King}{Zhao
  et~al\mbox{.}}{2014}]%
        {DBLP:conf/cikm/ZhaoMK14}
\bibfield{author}{\bibinfo{person}{Tong Zhao}, \bibinfo{person}{Julian~J.
  McAuley}, {and} \bibinfo{person}{Irwin King}.}
  \bibinfo{year}{2014}\natexlab{}.
\newblock \showarticletitle{Leveraging Social Connections to Improve
  Personalized Ranking for Collaborative Filtering}. In
  \bibinfo{booktitle}{\emph{CIKM'14}}.
\newblock


\bibitem[\protect\citeauthoryear{Zhong, Shah, Sundaravadivelan, and
  Sastry}{Zhong et~al\mbox{.}}{2013}]%
        {icwsm2013zhong}
\bibfield{author}{\bibinfo{person}{Changtao Zhong}, \bibinfo{person}{Sunil
  Shah}, \bibinfo{person}{Karthik Sundaravadivelan}, {and}
  \bibinfo{person}{Nishanth Sastry}.} \bibinfo{year}{2013}\natexlab{}.
\newblock \showarticletitle{Sharing the Loves: Understanding the How and Why of
  Online Content Curation}. In \bibinfo{booktitle}{\emph{ICWSM'13}}.
\newblock


\end{thebibliography}

\end{document}